\pdfoutput=1
\documentclass[11pt]{article}
\usepackage{amsmath,amssymb,graphicx,amsthm}
\usepackage{diagbox}
\usepackage{hyperref}
\usepackage[nosort]{cite}
\usepackage{mathtools}

\usepackage{multicol}

\usepackage{color}
\definecolor{darkred}{rgb}{0.65,0.15,0}
\definecolor{newgreen}{rgb}{0.2,0.62,0.14}
\definecolor{darkblue}{rgb}{0.1,0.15,0.7}
\definecolor{purple}{rgb}{0.35,0.1,0.55}
\hypersetup{pdfborder={0 0 0},colorlinks=true,urlcolor=blue,citecolor=blue,linkcolor=darkred,linktocpage=true}

\numberwithin{equation}{section}

\newcommand{\nn}{\nonumber}
\newcommand{\tA}{{\mathcal{A}}}
\newcommand{\tS}{{\mathcal{S}}}
\newcommand{\cX}{{\mathcal{X}}}
\newcommand{\cY}{{\mathcal{Y}}}
\newcommand{\tV}{{\mathfrak{V}}}
\newcommand{\tW}{{\mathfrak{W}}}

\newcommand{\cI}{{\mathcal{I}}}
\newcommand{\cJ}{{\mathcal{J}}}

\newcommand{\lb}{\left[}
\newcommand{\rb}{\right]}

\newcommand{\mf}[1]{{\mathfrak{#1}}}

\newcommand{\K}{{\rm k}_0}

\newcommand{\eprint}[1]{{\href{http://arxiv.org/abs/#1}{[\texttt{#1}]}}}
\newcommand{\eprintN}[1]{{\href{http://arxiv.org/abs/#1}{[\texttt{#1 [hep-th]}]}}}

\newcommand{\eprintRT}[1]{{\href{http://arxiv.org/abs/#1}{[\texttt{#1 [math.RT]}]}}}
\newcommand{\eprintGR}[1]{{\href{http://arxiv.org/abs/#1}{[\texttt{#1 [math.GR]}]}}}
\newcommand{\doi}[2]{{\hypersetup{urlcolor=darkred}\href{http://dx.doi.org/#2}{#1}\hypersetup{urlcolor=blue}}}

\newcommand{\al}{\alpha}

\newcommand{\bal}{\mathring\alpha}
\newcommand{\bbe}{\mathring\beta}
\newcommand{\ga}{\gamma}
\newcommand{\si}{\sigma}
\newcommand{\Ga}{\Gamma}
\newcommand{\tG}{{\tilde\Gamma}}
\newcommand{\de}{\delta}

\newcommand{\Ta}{{\tt a}}
\newcommand{\Tb}{{\tt b}}
\newcommand{\Tc}{{\tt c}}
\newcommand{\Td}{{\tt d}}

\newcommand{\ints}{\mathbb{Z}}

\newcommand{\rX}{{\rm X}}

\newcommand{\beq}{\begin{equation}}
\newcommand{\eeq}{\end{equation}}
\newcommand{\bea}{\begin{eqnarray}}
\newcommand{\eea}{\end{eqnarray}}

\begin{document}
\hypersetup{pageanchor=false}

\mbox{}
\vspace{15mm}

\begin{center}

{\LARGE \bf \sc  Generalised holonomies and $K(E_9)$}\\[5mm]

\vspace{6mm}
\normalsize
{\large  Axel Kleinschmidt${}^{1,2}$ and Hermann Nicolai${}^{1}$}

\vspace{10mm}
${}^1${\it Max-Planck-Institut f\"{u}r Gravitationsphysik (Albert-Einstein-Institut)\\
Am M\"{u}hlenberg 1, DE-14476 Potsdam, Germany}
\vskip 1 em
${}^2${\it International Solvay Institutes\\
ULB-Campus Plaine CP231, BE-1050 Brussels, Belgium}

\vspace{20mm}

\hrule

\vspace{5mm}

 \begin{tabular}{p{14cm}}
The involutory subalgebra $K(\mf{e}_9)$ of the affine Kac--Moody algebra $\mf{e}_9$ was recently shown to admit an infinite sequence of unfaithful representations of ever increasing 
dimensions~\cite{Kleinschmidt:2021agj}. We revisit these representations and describe their 
associated ideals in more detail, with particular emphasis on two chiral versions 
that can be constructed for each such representation. For every such unfaithful representation 
we show that the action of $K(\mf{e}_9)$ decomposes into a direct sum
of two mutually commuting (`chiral' 
and `anti-chiral') parabolic algebras with Levi subalgebra $\mf{so}(16)_+\,\oplus\,\mf{so}(16)_-$.
We also spell out the consistency conditions for uplifting such representations  to 
unfaithful representations of $K(\mf{e}_{10})$. 
From these results it is evident that the holonomy groups so far discussed in the literature 
are mere shadows (in a Platonic sense) of a much larger structure.
\end{tabular}

\vspace{6mm}
\hrule
\end{center}

\thispagestyle{empty}

\newpage
\setcounter{page}{1}
\hypersetup{pageanchor=true}

\setcounter{tocdepth}{2}
\tableofcontents

\bigskip

\section{Introduction} 

In studies of supersymmetric solutions of supergravity, a central role is played by the 
Killing spinor equation that expresses the vanishing of the supersymmetry variation of the gravitino 
\begin{align}
\label{eq:KSE}
\delta_\epsilon \Psi_M \,=\, \widehat{D}_M(\omega,F) \epsilon \,\equiv \,
 \left[ \partial_M + \frac14 \omega_{M\, AB} \Gamma^{AB} + F\cdot \Gamma \right] \epsilon = 0\,.
\end{align}
Here, $\epsilon$ represents the spin-$\frac12$ local supersymmetry parameter, $\omega_{M\,AB}$ the components of the spin connection and $F\cdot \Gamma$ is shorthand
for flux contributions, where the components of the field strengths are contracted 
with elements of the gamma matrix algebra; the explicit form of this term depends
on the theory in question. For maximal supergravity, the idea that the combination of 
the spin connection and the  flux terms should be interpreted as a generalised connection 
goes back to \cite{CJ,dWN}. More recently, and in a more general context,
it was pointed out that these flux terms extend 
the notion of holonomy of the spin bundle to that of a generalised holonomy~\cite{Duff:1990xz,Duff:2003ec,Hull:2003mf,Lu:2005im}. The latter has been a subject of intense study in connection with classifying supersymmetric solutions, also in the context of (exceptional) generalised geometry~\cite{Grana:2005jc,Grana:2005sn,Gabella:2009wu,Coimbra:2014uxa,Coimbra:2016ydd}. In particular, the groups 
\beq\label{oldGH}
SO(16)_+ \times SO(16)_-\,\subset \, SO(32) \, \subset \, SL(32,\mathbb{R})
\eeq 
have been put forward as candidate generalised holonomy groups of maximal supergravity in~\cite{Duff:2003ec,Hull:2003mf,West:2003fc}, and we shall focus on this theory in the following. 
The basis for this conjecture and the chain of embeddings in~\eqref{oldGH}
is that the $D=11$ gamma matrices appearing in~\eqref{eq:KSE} generate the whole 
Lie algebra of $\mf{sl}(32)$ upon commutation (together with its subalgebras
$\mf{so}(32)$ and $\mf{so}(16)\oplus \mf{so}(16)$), such that the $32$-component parameter 
$\epsilon$ can be viewed as belonging to the fundamental representation of this group. 

However, it was already pointed out in~\cite{Keurentjes:2003yu} that these conjectures are problematic at the group level in the sense that they do not contain the correct R-symmetry groups appearing upon dimensional reduction, notably the $Spin(16)$ symmetry in 
$D=3$ maximal supergravity. This incompatibility of the group structures negates the idea that the proposed generalised holonomy groups could be true symmetry groups of $D=11$ supergravity, although their properties continue to be useful for the study of supersymmetric solutions. A further argument against $SO(32)$ or $SL(32)$ as symmetries is that, while these groups can act on the spin-$\frac12$ supersymmetry parameter $\epsilon$, they cannot act on the propagating fermion of the theory, namely the gravitino: for maximal supergravity, the latter
is a vector spinor (`spin-$\frac32$') representation of the Lorentz group $Spin(1,10)$, 
but there is no way to turn it into a representation of $SL(32)$.

A resolution of these issues was proposed in the context of studies of Kac--Moody symmetries of supergravity. It was found in~\cite{deBuyl:2005zy,Damour:2005zs,deBuyl:2005sch,Damour:2006xu}
that the involutory subalgebra $K(\mf{e}_{10})$ of the Kac--Moody Lie algebra $\mf{e}_{10}$ admits a $32$-dimensional representation, dubbed spin-$\frac12$ for obvious reasons, as well as a $320$-component representation that corresponds to the spin-$\frac32$ gravitino. These representations have the property that they lift correctly to the spin cover $\widetilde{K(E_{10})}$ of the corresponding group (see~\cite{Harring:2019} for a discussion of the spin cover).
Moreover, the representations have the property that they possess the correct branching to the other maximal supergravity theories, including the chiral fermions of type IIB, which cannot be obtained by usual dimensional reduction~\cite{Kleinschmidt:2006tm}, and thus provide a common origin for both IIA and IIB fermions.
Since the fermionic representations are finite-dimensional representations of an infinite-dimensional group, they are unfaithful and there is thus a huge kernel of the representation map. The corresponding quotient groups turn out to be $SO(32)$ for the spin-$\frac12$ representation and the non-compact $SO(288,32)$ for the spin-$\frac32$ representation
\cite{KNV}.\footnote{For the spin-$\frac12$ and 
spin-$\tfrac32$ representations of $K(\mf{e}_{11})$ defined in~\cite{Kleinschmidt:2006tm} the corresponding quotients algebras are $\mf{sl}(32)$ and $\mf{sl}(352)$, respectively. See also~\cite{West:2003fc} and~\cite{Steele:2010tk} for work related to the spin-$\tfrac12$ and spin-$\tfrac32$ representations of $K(\mf{e}_{11})$.} The significance 
 of these results is that they represent incontrovertible evidence for the fact that \eqref{oldGH} 
are not the actual generalised holonomy groups, but that the chain of embeddings \eqref{oldGH}
must be replaced by \cite{NS}
\beq\label{newGH}
\widetilde{K(E_{9})}\,\subset \, \widetilde{K(E_{10})}\,\subset \,\widetilde{K(E_{11})}\,.
\eeq
Accordingly, the finite-dimensional symmetry groups which have appeared so far must be 
interpreted as quotient groups which are obtained by dividing $\widetilde{K(E_{n})}$ 
(for $n=9,10,11$) by the annihilator ideals of the corresponding unfaithful spinor representations.
Hence our claim that the holonomy groups identified so far are merely shadows of the full 
sequence expressed by \eqref{newGH} (and their analogs for non-maximal supergravities).
Since the actual group is $\widetilde{K(E_{n})}$, the topological obstructions found in~\cite{Keurentjes:2003yu} disappear, as it is no longer necessary to embed the R-symmetry $Spin(16)$ into $SL(32)$ but it embeds correctly (as a quotient) into $\widetilde{K(E_{n})}$. In this sense, the proposed Kac--Moody symmetries are much more promising candidates, see also~\cite{Kleinschmidt:2007zd} for a related discussion.

The purpose of this paper is to amplify this point, with special emphasis on the affine case,
and in this way actually {\em prove} part of the claim \eqref{newGH}. Since the 
finite-dimensional spin-$\frac12$ and spin-$\frac32$ representations furnish only infinitesimal glimpses of the huge Kac--Moody-type symmetry, it is important to understand the structure of its representations better. First steps were taken for $K(\mf{e}_{10})$ by the construction of two
new `higher spin' representations in~\cite{KN3,KN5} (see also \cite{KNV} for a review), 
where the `spin' actually refers to DeWitt superspace. In contrast to the spin-$\frac12$ and 
spin-$\frac32$ representations these can no longer be deduced from supergravity.
A more systematic study in the case of affine $\mf{e}_9$, relevant in the context of $D=2$ maximal supergravity~\cite{Julia,Julia2,Nicolai:1987kz,NS}, was initiated in~\cite{Kleinschmidt:2021agj} where a general construction of infinite families of unfaithful representations was given. This construction, reviewed in section~\ref{sec:rev} below, hinges on some properties that are specific to the affine case, but it is hoped that this serves as a stepping stone to the more interesting hyperbolic algebra $\mf{e}_{10}$. This hope stems from the fact that all known representations of $K(\mf{e}_{10})$ split into two chiral halves when decomposed under $K(\mf{e}_9)\subset K(\mf{e}_{10})$ and the resulting $K(\mf{e}_9)$ representations belong to the family of representations constructed in~\cite{Kleinschmidt:2021agj}. 
We stress that the converse is not true: not every $K(\mf{e}_9)$-representation can be `doubled' and uplifted to a $K(\mf{e}_{10})$ representation. It is one of the aims of this paper to find conditions when this is possible. 

As one of our main results we will exhibit for the affine case
the general structure of  the quotient group $Q$ of $\widetilde{K(E_{9})}$ as a product of two mutually commuting (`chiral' and `anti-chiral') parabolic groups 
(see \eqref{GH} for the Lie algebra version of this statement)
\beq\label{GHG}
Q = \big( Spin(16)_+ \ltimes U_+ \big) \times \big(Spin(16)_- \ltimes U_- \big)\,,
\eeq
where the Levi subgroups of the two parabolic groups are $Spin(16)_\pm$ and the unipotent subgroups $U_\pm$ are $N$-step unipotent, with $N$ referring to a truncation condition in the construction of the unfaithful representations. This is the infinite series of groups that
generalises the left-most entry in \eqref{oldGH}. The notion of chirality $\pm$ in the construction is related mathematically to choosing one of two fixed points of an involution but in terms of $D=2$ supergravity it corresponds to space-time chirality as we explain in more detail in section~\ref{sec:E10ext}. We also stress that depending on details of the construction discussed in section~\ref{sec:ideals}, some of the finite $\mathbb{Z}_2$ factors in the center of $Spin(16)$ can act trivially.
From~\eqref{GHG} we see that the proper 
generalisation of the holonomy group to non-trivial spinor representations therefore differs substantially from~\eqref{oldGH}. In particular there is no limit on the size of the group $Q$, since
the `cutoff parameter' $N$ can be taken arbitrarily large. For maximal supergravity
the structure of \eqref{GHG} is in complete accord with the known action of $\widetilde{K(E_9)}$ on the
supergravity fermions that was first exhibited in \cite{NS}.

Our motivation for this detailed study of representations of $K(\mf{e}_9)$ is twofold. Firstly, we can use the representations to obtain an ever more faithful description of the non-Kac--Moody structure $K(\mf{e}_9)$ by increasing the value of $N$.  Secondly, the proposal of $E_{10}$ symmetry in M-theory~\cite{DHN} is lacking a proper supersymmetric extension~\cite{Kleinschmidt:2014uwa}, a fact that can be possibly traced back to the imbalance between the faithful bosonic representation of $\mf{e}_{10}$ on the symmetric space $E_{10}/K(E_{10})$ and the unfaithful action of $K(\mf{e}_{10})$ on the supergravity fermions. If the present construction of increasingly faithful representations of $K(\mf{e}_9)$ leads to increasingly faithful representations of $K(\mf{e}_{10})$, this can shed new light on the de-emergence of space and its replacement by algebraic concepts as discussed in~\cite{Damour:2005zs}.

Since $K(\mf{e}_{10})$ differs from $K(\mf{e}_9)$ by exactly one simple Berman 
generator (designated by $x_1$ with our labeling of the $E_{10}$ Dynkin diagram, see figure~\ref{fig:e10dynk} in section~\ref{sec:E10ext}), 
representations of $K(\mf{e}_9)$ that can be doubled to $K(\mf{e}_{10})$ representations then must have a realisation of this Berman generator that mixes
and interchanges the two parabolic subgroups in~\eqref{GHG}. The relations that this Berman generator must satisfy are known and spelt out in section~\ref{sec:E10ext}. The problem of finding $K(\mf{e}_{10})$ representation is thus greatly reduced, but it is worth stressing that the action of this extra Berman generator is not simply an interchange of the two chiral halves. As we 
shall see with the concrete examples of the known $K(E_{10})$ representations,
the very existence of these representations relies on subtle consistency conditions 
that go beyond the affine construction in an essential way, 
and whose general form remains to be 
fully explored. We also note that the structure of \eqref{GHG} implies that $K(\mf{e}_{10})$ 
representations obtained in this way must necessarily have {\em non-compact} 
quotients whenever the unipotent parts $U_\pm$ are non-trivial. This explains one of 
the strangest features of our construction,
namely the fact that these quotient groups are generically non-compact 
even though they descend from a `maximally compact' subgroup \cite{KNV}.

 However, the main challenge at
this point concerns the extension of our results to $K(\mf{e}_{10})$, with the idea that its
corresponding unfaithful representations can be constructed by appropriately `gluing' two chiral
representations of  $K(\mf{e}_9)$, and by combining the two nilpotent subalgebras in \eqref{GH},
as we sketch in section~\ref{sec:E10ext}. The latter strategy is suggested by the fact that 
all known unfaithful spinorial representations of $K(\mf{e}_{10})$ are of this type.
In section~\ref{sec:E10ext} we also state the consistency conditions that must be satisfied in order
for such an `uplift' to $K(\mf{e}_{10})$ to work. This is important because we know that not 
every representation of $K(\mf{e}_{9})$ admits such an uplift to $K(\mf{e}_{10})$ (an example
is obtained by truncating a $\bf{16}\oplus \bf{16}$ from the spin-$\frac32$ representation). Spinor representations for involutory subalgebras of other indefinite Kac--Moody algebras have also been discussed in~\cite{Damour:2009zc,Hainke:2014,Lautenbacher:2017,KNV}.

Let us comment on the possible relevance of our results in view of some of the Kac--Moody conjectures in the literature. Recently, $E_n$ symmetries have been central in the context of exceptional field theory, see~\cite{Hohm:2019bba,Berman:2020tqn} for reviews, and representations of $\widetilde{K(E_n)}$ enter when discussing fermions~\cite{Godazgar:2014nqa,Baguet:2016jph,Bossard:2019ksx}, where in particular the last reference discusses the Kac--Moody case $n=11$. Exceptional field theories are duality \textit{covariant} formulations of families of maximally supersymmetric theories. Focussing on a single maximally supersymmetric theory breaks the $E_n$ symmetry but we expect the $\widetilde{K(E_n)}$ symmetry to remain intact, much in the same way as the reformulations of $D=11$ supergravity in~\cite{dWN,Nicolai:1986jk} maintain a larger local symmetry. 

By contrast, the conjectures of $E_{10}$~\cite{DHN} or $E_{11}$~\cite{West:2001as} symmetries of M-theory assert that the bosonic Kac--Moody symmetries are fully intact in (classical) M-theory, and at this point it is not clear what the fermionic extension of these conjectures will be. 
In particular, it is not clear whether these theories are merely supposed to be (albeit very complicated) 
reformulations of maximal supergravities with essentially the same local dynamics, or whether
the dynamics involve local degrees of freedom beyond the ones of maximal supergravity. 
The former point of  view has been adopted in much of the literature on exceptional  field theory. The second point of 
view is supported by  the fact that there do exist theories `beyond' string theory, and with different 
(non-perturbative) dynamics, namely the $D=11$ supermembrane \cite{Bergshoeff:1987cm}
and the associated maximally supersymmetric $SU(\infty)$ matrix 
theory~\cite{deWit:1988wri}. We hope that the results presented in 
this paper can contribute to the clarification of these issues, and we expect that in either case
further advances will make recourse to some of the structures derived in this paper.

The results presented in this paper are given for $K(\mf{e}_9)$, but most of the general structure is applicable to any $K(\mf{g})$ where $\mf{g}$ is a non-twisted affine algebra. One possible other arena for the considerations of this paper would therefore be pure $\mathcal{N}=1$ supergravity reduced from $D=4$ to $D=2$ dimensions where the (Geroch) symmetry algebra $\mf{g}$ is affinised $\mf{sl}(2)$~\cite{Breitenlohner:1986um}.

The structure of this article is as follows. We first review the general construction of $K(\mf{e}_9)$ representations of~\cite{Kleinschmidt:2021agj} in section~\ref{sec:rev}. In section~\ref{sec:ideals}, we discuss the structure of the associated ideals in more detail and in particular introduce the notion of chirality, showing that the chiral quotients commute. Section~\ref{sec:rtbas} connects the results of this paper to the explicit higher-spin representations of $K(\mf{e}_{10})$ found in~\cite{KN3} and how they reduce to $K(\mf{e}_9)$. In section~\ref{sec:E10ext} we then discuss the converse problem and discuss conditions for lifting $K(\mf{e}_9)$ representations to $K(\mf{e}_{10})$. 

\subsubsection*{Acknowledgements}

We are grateful to Ralf K\"ohl and Robin Lautenbacher for discussions. The work of  H.N. has received funding from the European Research 
 Council (ERC) under the  European Union's Horizon 2020 research and 
 innovation programme (grant agreement No 740209).

\section{Parabolic and filtered algebras}
\label{sec:rev}

In this section, we review the basic set-up from~\cite{Kleinschmidt:2021agj}. 

\subsection{\texorpdfstring{$K(E_9)$ from $E_9$}{K(E9) from E9}}

The Kac--Moody algebra $\mf{e}_9$ can be defined in terms of a Chevalley--Serre basis and associated relations described by its Dynkin diagram. As for any affine Lie algebra it can be equivalently given in terms of a centrally extended loop algebra as
\begin{align}
\mf{e}_9 = \mf{e}_8  [t,t^{-1}]\oplus \mathbb{R} k \oplus \mathbb{R} d
\end{align}
where $\mf{e}_8[t,t^{-1}] \cong \mathbb{R}[t,t^{-1}] \otimes \mf{e}_8$ denotes $\mf{e}_8$-valued Laurent polynomials in a formal variable $t$ and $k$ and $d$ are semi-simple elements of the algebra, with $k$ being central. The commutation relations are given by
\begin{align}
\label{eq:CRaff}
\left[ t^m \otimes x , t^n \otimes y \right] &= t^{m+n} \otimes [x,y] + m \delta_{m,n} (x,y) k\,,\nn\\
\left[ t^m \otimes x,k\right] = \left[d,k\right]&=0\,,\hspace{15mm} \left[t^m\otimes x, d\right] = m t^m\otimes x
\end{align}
for $m,n\in\mathbb{Z}$ and $x,y\in\mf{e}_8$ and with $(x,y)$ the canonically normalised Killing pairing on $\mf{e}_8$. 
We only consider the split real form of the algebra.

The $248$-dimensional algebra $\mf{e}_8$ has a Cartan--Chevalley involution $\mathring{\omega}$ whose fixed point set is the $120$-dimensional compact subalgebra $\mf{so}(16)$ of $\mf{e}_8$. The remaining $128$ generators transform in a spinor representation of $\mf{so}(16)$. We shall denote the generators in the associated Cartan decomposition as $(X^{IJ}, Y^A)$ with $X^{IJ}=X^{[IJ]}$ the $120$ $\mf{so}(16)$ generators while $Y^A$ are the $128$ coset generators. The commutation relations are
\begin{align}
\left[ X^{IJ} , X^{KL} \right] &= 2 \delta^{K[J} X^{I]L} - 2 \delta^{L[J} X^{I]K}\,,\nn\\
\left[ X^{IJ}, Y^A \right] &= -\frac12 \Gamma^{IJ}_{AB} Y^B \,,\\
\left[ Y^A, Y^B\right] &= \frac14 \Gamma^{IJ}_{AB} X^{IJ}\,. \nn
\end{align}
Indices can be raised and lowered with the $\mf{so}(16)$-invariant metrics $\delta_{IJ}$ for fundamental indices and $\delta_{AB}$ for spinor indices. The matrices $\Gamma^{IJ}_{AB}$ are taken from the $\mf{so}(16)$ Clifford algebra and they are of size $128{\times}128$ for fixed $I,J$. Conjugate spinor representations of $\mf{so}(16)$ will be denoted with indices $\dot{A}$. The involution $\mathring{\omega}$ acts on these basis generators by
\begin{align}
\mathring{\omega}(X^{IJ}) = X^{IJ} \,,\hspace{15mm}
\mathring{\omega}(Y^A) = -Y^A\,.
\end{align}
The Cartan--Chevalley involution $\mathring{\omega}$ on $\mf{e}_8$ can be extended to an involution $\omega$ on $\mf{e}_9$ by letting
\begin{align}
\label{eq:omE9}
\omega( t^n \otimes x) = t^{-n}\otimes \mathring{\omega}(x)\,,\quad\quad
\omega(d)=-d\,,\quad \omega(k) = - k \,.
\end{align}
We denote the invariant subalgebra under this involution by $K(\mf{e}_9)\equiv \mf{k}$. 

\subsection{Filtered and parabolic algebras}

A basis for $K(\mf{e}_9)$ can be given explicitly by 
\begin{subequations}
\label{eq:filtbas}
\begin{align}
\cX_n^{IJ} &\coloneqq \frac12 (t^n+t^{-n}) \otimes X^{IJ} &&\text{for $n\geq 0$,}\\
\cY_n^A &\coloneqq \frac12 (t^n-t^{-n}) \otimes Y^A &&\text{for $n>0$}\,.
\end{align}
\end{subequations}
The elements $d$ and $k$ are projected out when descending from $\mf{e}_9$ to $K(\mf{e}_9)$. 

We shall refer to the elements~\eqref{eq:filtbas} as the \textit{filtered basis} of $K(\mf{e}_9)$. The reason for this terminology is that the commutation relations in this basis are no longer graded (as in~\eqref{eq:CRaff}) but read
\begin{subequations}
\begin{align}
\lb \cX_m^{IJ} , \cX_n^{KL} \rb &= 2 \delta^{[I[K} \left( \cX_{m+n}^{L]J]} + \cX_{|m-n|}^{L]J]} \right)\, ,\\
\lb\cX_m^{IJ} , \cY_n^A \rb &= -\frac14 \Gamma^{IJ}_{AB} \left( \cY_{m+n}^B -\text{sgn}(m-n)\cY_{|m-n|}^B\right)\,,\\
\lb \cY_m^A ,\cY_n^B \rb &= \frac18 \Gamma^{IJ}_{AB} \left( \cX_{m+n}^{IJ} - \cX_{|m-n|}^{IJ}\right)\,.
\end{align}
\end{subequations}

As the involution~\eqref{eq:omE9} sends $t \to t^{-1}$ one can also consider a different algebra that is more adapted to the fixed points of this involution on Laurent polynomials $\mathbb{R}[t,t^{-1}]$. The fixed points are given by $t_*=\pm 1$ and we consider the explicit change of variables
\begin{align}
\label{eq:chvar}
u = \frac{1\mp t}{1\pm t} \quad\Longleftrightarrow \quad
t = \pm \frac{1-u}{1+u}
\end{align}
for the two choices of fixed points. Around these fixed points we define the Lie algebra
\begin{align}
\label{eq:par}
\mathfrak{N}= \mathbb{R}[[u^2]] \otimes \langle X^{IJ}\rangle \oplus u\mathbb{R}[[u^2]] \otimes \langle Y^A\rangle
\end{align}
with the bracket defined in the obvious way and allowing for arbitrary formal power series. Written explicitly, a basis is of this Lie algebra is given by
\begin{align}
\tA^{IJ}_{2n} &\coloneqq u^{2n} \otimes X^{IJ}  &&\text{for $n\geq 0$,}\nn\\
\tS^A_{2n+1} &\coloneqq u^{2n+1} \otimes Y^A  &&\text{for $n\geq 0$.}
\end{align}
We stress that the Lie algebra~\eqref{eq:par} is allowed to contain arbitrary formal power series in $u$. Since only positive powers of $u$ arises, the commutation relations of the basis elements are graded according to
\begin{subequations}
\label{eq:ke9rels}
\begin{align}
\lb \tA_{2m}^{IJ} , \tA_{2n}^{KL} \rb &= \frac12 \delta^{JK} \tA_{2(m+n)}^{IL}- \frac12 \delta^{IK} \tA_{2(m+n)}^{JL}- \frac12 \delta^{JL} \tA_{2(m+n)}^{IK}+ \frac12 \delta^{IL} \tA_{2(m+n)}^{JK}
\,,\\[2mm]
\lb \tA_{2m}^{IJ} , \tS_{2n+1}^{KL} \rb &= -\frac14\Gamma^{IJ}_{AB} \tS_{2(m+n)+1}^{B}\,,\\[2mm]
\label{ke9rels2c}
\lb \tS_{2m+1}^A, \tS_{2n+1}^B \rb &= \frac18 \Gamma^{IJ}_{AB} \tA_{2(m+n+1)}^{IJ}\,.
\end{align}
\end{subequations}
We refer to this algebra as a \textit{parabolic} Lie algebra as the Levi part $\mf{so}(16)$, generated by $\tA_0^{IJ}$, acts on the (generalised) nilpotent part with positive subscripts. 

One of the main results of~\cite{Kleinschmidt:2021agj} was that there are injective Lie algebra homomorphisms $\rho_{\pm} : K(\mf{e}_9) \to \mathfrak{N}$. These can be obtained by expanding out the explicit change of variables~\eqref{eq:chvar} according to
\begin{align}
t^n + t^{-n} = (\pm 1)^n \sum_{k\geq 0} a_{2k}^{(n)} u^{2k} \,,\quad
t^n- t^{-n}  = (\pm 1)^n \sum_{k\geq 0} a_{2k+1}^{(n)} u^{2k+1}
\end{align}
for $n\geq 0$ and $n>0$, respectively. The two choices of sign correspond to the two choices in~\eqref{eq:chvar}. The two Lie algebra homomorphisms are then defined explicitly by
\begin{align}\label{rhoXY}
\rho_{\pm} (\cX_n^{IJ}) = (\pm 1)^n \frac12  \sum_{k\geq 0} a_{2k}^{(n)} \tA_{2k}^{IJ} \,,\quad
\rho_{\pm} (\cY_n^A) = (\pm 1)^n \frac12 \sum_{k\geq 0} a_{2k+1}^{(n)} \tS_{2k+1}^A\,.
\end{align}
As explained in~\cite{Kleinschmidt:2021agj}, the homomorphisms are not surjective as one would require power series in $\mathbb{R}[[t,t^{-1}]]$ that, however, do not behave well under multiplication. 

Explicit formulas for the coefficients are 
\begin{subequations}
\begin{align}
a_{2k}^{(n)} &= 2 \sum_{\ell=0}^n \binom{2n}{2\ell} \binom{k-\ell+n-1}{k-\ell} &&\text{for $n\geq 1$ and $k\geq 1$}
\,,\\
a_{2k+1}^{(n)} &= -2 \sum_{\ell=0}^{n-1} \binom{2n}{2\ell+1}  \binom{k-\ell+n-1}{k-\ell}&&\text{for $n\geq 0$ and $k\geq 0$}
\,.
\end{align}
\end{subequations}
Also $a_{2k}^{(0)}=0$ for all $k\geq 1$ and $a_0^{(n)} = 2$ for all $n\geq 0$.

\subsection{Representations from truncations}

It is easy to construct finite-dimensional representations for the parabolic algebra $\mathfrak{N}$ defined in~\eqref{eq:par} by considering quotients of the algebra. Examples can be obtained by quotienting the ring of power series $\mathbb{R}[[u]]$ by the ideal $u^{N+1}\mathbb{R}[[u]]$ of power series whose lowest order term is $u^{N+1}$. As a vector space, the quotient corresponds to polynomials of degree at most $N$, but the quotient construction also endows the vector space with a product structure. This is given by working modulo terms of order $\mathcal{O}(u^{N+1})$. 
In terms of the parabolic generators this amounts to setting
\begin{equation}\label{cutoff}
\tA^{IJ}_{2k} = 0 \;\; \mbox{for $k > \lfloor N/2\rfloor$}\quad, \qquad
\tS_{2k+1}^A = 0 \;\; \mbox{for $k >\lfloor (N-1)/2\rfloor$}\,.
\end{equation}
There are two cases to be distinguished here, according to whether the highest `active'
generator is of type $\tA_{2k}$ or of type $\tS_{2k+1}$. Since these two cases are 
largely analogous we will for definiteness assume $N =2\, \K$ even and
\beq
\tA^{IJ}_{2k} = \tS^A_{2k-1} = 0  \qquad \mbox{for $k > \K= \frac{N}{2}$}
\eeq
in the remainder, such that the highest `active' generators are $\tA_{2\K}^{IJ}$ and
$\tS_{2\K -1}^A$.
This quotient on the ring of power series induces a quotient Lie algebra of $\mathfrak{N}$ that we denote by $\mathfrak{N}_N$. Representations of this quotient Lie algebra can be constructed by considering all elements of degree at most $N$ in the universal enveloping algebra of $\mathfrak{N}_N$ acting on a given $\mf{so}(16)$ module $\tV_0$. In practice, this means that we are considering the graded representation
\begin{align}
\tW_N = \bigoplus_{i=0}^N \tV_i
\end{align}
with the constituent $\mf{so}(16)$ modules (with $N=2\K$)
\begin{align}\label{tWN}
\tV_0 &\nn\\
\tV_1 &= \langle \tS_1^A \tV_0 \rangle\nn\\
\tV_2 &= \langle \tS_1^{(A} \tS_1^{B)} \tV_0\rangle \oplus \langle \tA_2^{IJ} \tV_0\rangle\\
\vdots\nn \\
\tV_N &= \langle \tS_1^{(A_1} \cdots \tS_1^{A_N)} \tV_0 \rangle  \oplus \, \cdots \,\oplus
\langle\tS_1^A \tS_{N-1}^B \tV_0\rangle \oplus  \langle \tA^{IJ}_N \tV_0 \rangle
\end{align}
where the generators must be ordered in accordance with the Poincar\'e--Birkhoff--Witt theorem.
As these are representations of a quotient $\mathfrak{N}_N$ of the parabolic Lie algebra $\mathfrak{N}$ into which $K(\mf{e}_9)$ embeds injectively, the representations can be pulled back to finite-dimensional representations of $K(\mf{e}_9)$ using the homomorphisms $\rho_\pm$~\cite{Kleinschmidt:2021agj}. They provide a plethora of new representations.

Yet more representations can be obtained from \eqref{tWN} by further truncating 
away subrepresentations at any level together with their associated submodules of
$\tW_N$, as was already explained in \cite{Kleinschmidt:2021agj}. Unlike the quotient to $\tW_N$, removing further $\mf{so}(16)$ representations within one of the constituent $\tV_i$ may require a careful check that the resulting space is a representation of $K(\mf{e}_9)$. 
Nevertheless, the known examples of $K(\mf{e}_{10})$ representations
suggest that such extra truncations may be necessary for the existence of the over-extended
Berman generator $x_1$ and for a consistent uplift to $K(E_{10})$~\cite{KN5}.

\section{Ideals}
\label{sec:ideals}

The representations $\tW_N$ defined in the previous section are characterised by the fact 
that $\tA_{2k}^{IJ}=0$ and $\tS_{2k-1}^A=0$ for $k>\K$, where we fix $\K$
throughout this section. The case when the highest active generator is of type
$\tS$ can be dealt with similarly.

We therefore have the finite sum relations valid in these representations
\begin{subequations}
\label{eq:Ktrunc}
\begin{align}
\rho_\pm (\cX_n^{IJ} ) = (\pm 1)^n \frac12 \sum_{k=0}^{\K} a_{2k}^{(n)} \tA_{2k}^{IJ}\,,\\
\rho_\pm (\cY_n^A ) = (\pm 1)^n \frac12 \sum_{k=1}^{\K} a_{2k-1}^{(n)} \tS_{2k-1}^A\,.
\end{align}
\end{subequations}
We can consider the first relation for $0\leq n \leq \K$ and the second one for $1\leq n\leq \K$.
From the explicit form of the $a_{2k}^{(n)}$ one can check that these represent linear systems of relations with unique solutions for $\tA_{2k}$ and $\tS_{2k-1}$ in terms of $\cX_n$ and
$\cY_n$ for $n\leq \K$. Substituting this solution with the $\tA_{2k}^{IJ}$ and $\tS_{2k+1}^A$ 
in terms of the corresponding $\cX_n^{IJ}$ and $\cY^A_n$ back into \eqref{eq:Ktrunc}
we can express any of the higher $\cX_n^{IJ}$ and $\cY_n^A$ for $n >\K$
in terms of these. Explicitly, we have \textit{in the representation}
\begin{subequations}
\label{eq:ideal1}
\begin{align}
(\pm 1)^n \cX_{n}^{IJ} &=  \sum_{m= 0}^{\K} ( \pm 1)^m c_{n,m} \cX_m^{IJ}\,,\\
(\pm 1)^n \cY_n^A &=  \sum_{m = 1}^{\K} (\pm 1)^m d_{n,m} \cY_m^A
\end{align}
\end{subequations}
where the coefficients are given by 
\begin{align}
\label{eq:coeffs}
c_{n,m} \equiv c_{n,m}^{(\K)} 
 := \prod_{\substack{0\leq j\leq \K \\ j\neq m}} \frac{n^2-j^2}{m^2-j^2} \,,\hspace{10mm}
d_{n,m} \equiv d_{n,m}^{(\K)} 
 := \frac{n}{m} \prod_{\substack{1\leq j\leq \K \\ j\neq m}}  \frac{n^2-j^2}{m^2-j^2} \,.
\end{align}
where always $m\leq \K$ (below we will often suppress the superscript $\K$).
For $0\leq n\leq \K$, the relations~\eqref{eq:ideal1} 
become tautologies, as in this range $c^{(\K)}_{n,m} =d^{(\K)}_{n,m} = \delta_{n,m}$
(remember that the second index $m$ is always restricted to a finite range).
Consequently, only the $\cX_n^{IJ}$ for $0\leq n\leq \K$ and the $\cY_n^A$ for
$1\leq n\leq \K$ are independent, whereas the higher index objects depend linearly on
them via the above ideal relations. Put differently, for $\cX_n^{IJ}$ with $n>\K$ 
and $\cY_n^A$ with $n>\K$ the relations~\eqref{eq:ideal1} are non-trivial relations among the representation matrices of $K(\mf{e}_9)$. Importantly, these are valid \textit{only} in the representations 
obtained by truncating the parabolic algebra.  For ease of notation we omit the homomorphisms $\rho_\pm$ and representation maps from the formul\ae{} that are always to be understood in the representation $\tW_N$.

The above coefficients are the unique solutions to the problem of finding coefficients that satisfy the conditions that $c^{(\K)}_{n,m} =d^{(\K)}_{n,m} = \delta_{n,m}$ for small $0\leq n\leq \K$,
and are polynomial of the right degree and parity in $n$.

\subsection{Chiral ideals}

We denote representations where the relations~\eqref{eq:ideal1} are satisfied with a definite choice of sign as \textit{chiral} representations, and write them as $S_\pm$.  In the following discussion we shall not write out the $\mf{so}(16)$ indices $IJ$ and $A$ on $\cX_m^{IJ}$ and $\cY_m^A$ as they are spectators in the whole discussion.
Corresponding to the two choices of sign in~\eqref{eq:ideal1} 
we define the following elements 
\begin{align}
\label{eq:idgen}
\cI_n^\pm &\coloneqq \cX_n -\sum_{m=0}^{\K} (\pm 1)^{n+m} c_{n,m} \cX_m \,,\nn\\
\cJ_n^\pm &\coloneqq \cY_n -\sum_{m=1}^{\K} (\pm 1)^{n+m} d_{n,m} \cY_m \,.
\end{align}
The $\cI_n^\pm$ and the $\cJ_n^\pm$ are only non-zero for $n>\K$. 
These are represented trivially in the truncated representations for any $n$,
and therefore we have the two ideals
\begin{equation}
\label{eq:ipm}
\mf{i}_\pm \,=\,   \bigoplus_{n> \K} \mathbb{R}\, \cI_n^\pm \,\oplus\, 
 \bigoplus_{n> \K} \mathbb{R}\, \cJ_n^\pm  \,\subset \, \mf{k}\,.
\end{equation}
The non-direct sum $\mf{i}_+ + \mf{i}_-\subset \mf{k}$ is consequently spanned by the even and odd 
combinations 
\begin{align}\label{ieven}
 \cX_{2n} - \sum_{m=0}^{\lfloor \frac{\K}2\rfloor } 
c_{2n,2m}^{(\K)} \cX_{2m}    \quad
& \text{and} \quad
\cY_{2n} - \sum_{m=1}^{\lfloor \frac{\K}2\rfloor } d_{2n,2m}^{(\K)} \cY_{2m} \,, \nonumber\\[2mm]
 \cX_{2n+1} - \sum_{m=0}^{\lfloor \frac{\K-1}2\rfloor } c_{2n+1,2m+1}^{(\K)} \cX_{2m+1} \quad
&\text{and} \quad
\cY_{2n+1} - \sum_{m=0}^{\lfloor \frac{\K-1}2\rfloor } d_{2n+1,2m+1}^{(\K)} \cY_{2m+1}  \,,
\end{align} 
and the combinations
\begin{align}
\label{ifinite}
\sum_{m=0}^{ \lfloor \frac{\K}2\rfloor} c^{(\K)}_{2n+1,2m} \cX_{2m}  \quad & \text{and} \qquad
\sum_{m=1}^{\lfloor \frac{\K}2\rfloor} d^{(\K)}_{2n+1,2m} \cY_{2m} \,, \nonumber\\[2mm]
\sum_{m=0}^{ \lfloor \frac{\K-1}2\rfloor} c^{(\K)}_{2n,2m+1} \cX_{2m+1}  \quad & \text{and} \qquad
\sum_{m=0}^{\lfloor \frac{\K-1}2\rfloor} d^{(\K)}_{2n,2m+1} \cY_{2m+1} \,.
\end{align}
These combinations are, respectively, obtained by adding and subtracting the
above ideal components for even and for odd $n$.

We can now show that these combinations together span all of $\mf{k}=K(\mf{e}_9)$.

\medskip
\noindent
{\bf Lemma.} We have
\begin{equation}
\label{eq:kipm}
\mf{i}_+ + \mf{i}_- \,=\, \mf{k} \,.
\end{equation}

\noindent
{\bf Proof:}  We first note that the relations \eqref{ifinite} only involve the generators $\cX_m$ 
and $\cY_m$ for $m\leq \K$. Consequently we have an {\em infinite 
set of relations} for finitely many objects, and these relations are not of finite corank due to the structure of the coefficients~\eqref{eq:coeffs}.
From this we conclude that all $\cX_m$ for $0\leq m\leq \K$ and all $\cY_m$ for 
$1\leq m \leq \K$ can be obtained by suitable linear combinations, and are
therefore contained in $\mf{i}_++\mf{i}_-$. Taking linear combinations with~\eqref{ieven} we then see that all $\cX_m$ and $\cY_m$ for any index are contained in $\mf{i}_++\mf{i}_-$ which therefore equals $\mf{k}$. \hfill  \qedsymbol

\subsection{Commuting chiral quotients}

We now consider the representation of $\mf{k}$ that is given by the direct sum $S_+\oplus S_-$ of two chiral representations $S_\pm$ with ideals $\mf{i}_\pm$ and associated quotient algebras 
\begin{align}
\mf{q}_\pm = \mf{k}/\mf{i}_\pm\,.
\end{align}
Note that the ideals do not fix the representations and the following statements should be correct 
for any choice of representations with the same ideal. The representations $S_+$ and $S_-$
are related to one another by exchanging $\cX_m \leftrightarrow (\pm 1)^m \cX_m$ and 
$\cY_m \leftrightarrow (\pm 1)^m \cY_m$, and in this sense equivalent.
From the explicit form of the ideals we see that representatives of the quotient algebras can be given solely in terms of the generators $\cX_0,\cX_1,\ldots, \cX_{\K}$ and $\cY_1,\cY_2,\ldots, \cY_{\K}$.

The ideal of the representation $S_+\oplus S_-$ is given by the intersection of ideals $\mf{i}_+ \cap \mf{i}_-$
and the quotient algebra therefore by
\begin{equation}
\label{eq:ii}
\mf{q} \,:=\, \mf{k}/ \mf{i} \,, 
\hspace{10mm} \mf{i}\coloneqq \mf{i}_+\cap \mf{i}_-\,.
\end{equation}

We first describe the intersection ideal $\mf{i}$ in more detail. This intersection consists of linear combinations of elements in $\mf{i}_\pm$ given in~\eqref{eq:ipm} that can be formed such that the choice of sign disappears. We discuss this for the case when $\K$ is even for concreteness; other cases can be analysed similarly. 

The spanning elements~\eqref{eq:idgen} 
have different signs depending on the chirality and the $\cI_n^\pm$ (resp. $\cJ_n^\pm$) therefore do not lie in the intersection ideal $\mf{i}$. We need to form linear combinations that are common to $\mf{i}_+$ and $\mf{i}_-$ where these alternating signs do not appear and we describe this in detail for the $\cX_m$. This can be done by choosing $\K+1$ appropriate elements $\cI_n^\pm$ that we take to be $\cI^\pm_{\K+1},\, \cI_{\K+2}^\pm, \ldots, \cI_{2\K+1}^\pm$. The $(\pm 1)^m \cX_m$ with  $m\leq \K$ can be expressed in terms of these elements so that we have found expressions that do not involve any alternating sign and these elements belong to the intersection ideal. Because they involve combinations of evenly indexed $\cX_m$ with $0\leq m\leq 2\K+1$, the range of indices
is extended over twice the previous range. We find that the intersection ideal $\mf{i}=\mf{i}_+\cap \mf{i}_-$is given by
\begin{align}
\label{eq:intid}
\mf{i} 
&= \bigoplus_{n>\K} \mathbb{R} \left( \cX_{2n} - \sum_{m=0}^{\K} c_{n,m}^{(\K)} \cX_{2m}\right) \oplus
 \bigoplus_{n>\K+1} \mathbb{R} \left( \cY_{2n} - \sum_{m=1}^{\K} d_{n,m}^{(\K)} \cY_{2m}\right)\nn\\
 &\quad \oplus  \bigoplus_{n>\K} \mathbb{R} \left( \cX_{2n+1} - \sum_{m=0}^{\K} e_{n,m}^{(\K)} \cX_{2m+1}\right) \oplus
  \bigoplus_{n>\K-1} \mathbb{R} \left( \cY_{2n+1} - \sum_{m=0}^{\K-1} f_{n,m}^{(\K)} \cY_{2m+1}\right)
\end{align}
with the coefficients from~\eqref{eq:coeffs} as well as
\begin{align}
e_{n,m}^{(\K)} = \prod_{\substack{j=-\K\\j\notin\{-m,m+1\}}}^{\K+1} \frac{n+j}{m+j}\,,
\quad\quad
f_{n,m}^{(\K)} = \frac{2n+1}{2m+1}\prod_{\substack{j=-\K+1\\j\notin\{-m,m+1\}}}^{\K} \frac{n+j}{m+j}\,.
\end{align}
Note that in~\eqref{eq:intid} the mode number $m$ on the coefficient and the one on the generator are not identical in contrast to~\eqref{eq:idgen}. A consistency check on these coefficients is that they must be polynomials of degree $2\K$ (for $\cX$) or $2\K-1$ (for $\cY$) with the requirement of being even/odd for $\cX_{2n}$ and $\cY_{2n}$ under $n\to -n$ as well as being even/odd for $\cX_{2n+1}$ and $\cY_{2n+1}$ under $n\to -n-1$. At the same time they must vanish for $n\leq \K (\pm1)$ as given by the summation ranges in~\eqref{eq:intid}.

We conclude from~\eqref{eq:intid} that the quotient algebra $\mf{q}$ defined in~\eqref{eq:ii} has representatives given in terms of $\cX_0,\cX_1,\ldots \cX_{2\K+1}$ and $\cY_1,\cY_2,\ldots , \cY_{2\K}$, exactly twice as many as the chiral quotients~$\mf{q}_{\pm}$.

Let us exemplify the procedure of obtaining~\eqref{eq:intid} for the case $\K=2$. The $\cI_n^{\pm}$ elements occurring in $\mf{i}_{\pm}$ are 
\begin{align}
\label{eq:K2ex}
\cI_{2n}^\pm &= \cX_{2n} - \frac14  (4n^2-1)(4n^2-4) \cX_0  \pm \frac43   n^2 (4n^2-4) \cX_1 -\frac1{3} n^2 (4n^2-1) \cX_2\,,\nn\\
\cI_{2n+1}^\pm &= \cX_{2n+1} \mp   (n^2+n)(4n^2+4n-3) \cX_0 + \frac13   (2n+1)^2 (4n^2+4n-3) \cX_1\nn\\
&\quad \mp \frac13(n^2+n) (2n+1)^2 \cX_2\,.
\end{align} 
For $\cI_n^\pm$ with $n\in\{0,1,2\}$ they vanish identically.  For other values of $n$ we can use this to solve for $\pm \cX_1$, $\mp \cX_0$ and $\mp \cX_2$. For example, as $\cI_4^\pm$ is given by
\begin{align}
\cI_4^\pm &= \cX_4 - 45 \cX_0 \pm 64 \cX_1 -20 \cX_2\,,
\end{align}
we have
\begin{align}
\pm \cX_1 &= \frac1{64} \left( \cI_4^\pm +45 \cX_0 +20\cX_2-\cX_4\right)\,.
\end{align}
Substituting this back into~\eqref{eq:K2ex} leads to expressions that are identical in $\mf{i}_+$ and $\mf{i}_-$. For the even case we get 
\begin{align}
\label{eq:I2ninter}
\cI_{2n}^\pm -  \frac1{48}   n^2 (4n^2-4)  \cI_4^\pm&= \cX_{2n} - \frac14  (n^2-1)(n^2-4) \cX_0  +\frac1{3} n^2 (n^2-4) \cX_2 - \frac1{12}   n^2 (n^2-1)\cX_4
\end{align}
as elements of $\mf{i}_\pm$. As the right-hand side does not depend on the choice of chirality, these elements belong to the intersection ideal. Among the even $\cX_n$ we therefore remain with $\cX_0$, $\cX_2$ and $\cX_4$ in the quotient algebra~\eqref{eq:ii}. Similarly, for the odd indices, we consider 
\begin{align}
\cI_3^\pm &= \cX_{3} \mp 10 \cX_0 +15 \cX_1 \mp 6 \cX_2\,,\nn\\
\cI_5^\pm &= \cX_{5} \mp 126 \cX_0 +155 \cX_1 \mp 50 \cX_2\,,
\end{align}
which can be solve to give
\begin{align}
\pm \cX_0 &= \frac1{128}\left( 25 \cI_3^\pm- 3 \cI_5^\pm +150\cX_! -25 \cX_3-3\cX_5\right)\,,\nn\\
\pm \cX_2 &= \frac1{128}\left(-63 \cI_3^\pm +5\cI_5^\pm +70\cX_1 +63 \cX_3 -5\cX_5\right)\,.
\end{align}
Substituting this back into~\eqref{eq:K2ex} leads to
\begin{align}
\label{eq:I2+1ninter}
&\quad \cI_{2n+1}^\pm +\frac18 (n{-}2)n (n{+}1)(n{+}3) \cI_3^\pm -\frac1{24} (n{-}1)n(n{+}1)(n{+}2) \cI_5^\pm \\
&=\cX_{2n+1} -\frac1{12} (n{-}2)(n{-}1)(n{+}2)(n{+}3)\cX_1 +\frac18(n{-}2)n(n{+}1)(n{+}3) \cX_3 -\frac1{24} (n{-}1)n(n{+}1)(n{+}2)\cX_5\,.\nn
\end{align}
Together with~\eqref{eq:I2ninter} we conclude that $\mf{q}$ includes the representatives $\cX_0,\cX_1,\ldots, \cX_5$. As a consistency check, we can see that the relations~\eqref{eq:I2ninter} and~\eqref{eq:I2+1ninter}  do not impose any constraint on these elements.
Performing the same analysis for the $\cY_n$ elements adds $\cY_1, \cY_2,\ldots , \cY_4$ as representatives to $\mf{q}$. 
(The case when the highest active generator is of $\tS$-type can be treated analogously.)

\medskip
We next discuss the relation between the quotients $\mf{q}_\pm$ and $\mf{q}$ of $\mf{k}$. Since we have trivially $\mf{i}\subset \mf{i}_\pm$ as an ideal, we can view $\mf{q}_\pm$ also canonically as quotients of $\mf{q}$ via
\begin{align}
\mf{q}_\pm \cong \mf{q} / \big( \mf{i}_\pm / \mf{i} \big)\,.
\end{align}
From the truncation construction it also follows that $\mf{q}_\pm$ are actually subalgebras of $\mf{q}$.
Using the relation~\eqref{eq:kipm} above we also obtain
\begin{align}
\label{eq:qpm2}
\mf{q}_\pm = \big(\mf{i}_+ + \mf{i}_-\big) / \mf{i}_\pm = \mf{i}_\mp / \mf{i}\,.
\end{align}

\medskip
\noindent
{\bf Proposition.} For any finite-dimensional representation of $\mf{k}$ defined 
by the cutoff conditions (\ref{cutoff}) the quotient algebra $\mf{q}$ decomposes 
into a direct sum of two mutually commuting subalgebras
\begin{equation}
\mf{q} \,=\, \mf{q}_+ \oplus \mf{q}_- \quad , \qquad  \big[ \mf{q}_+ , \mf{q}_- \big] = 0
\quad \text{as subalgebras of $\mf{q}$.}
\end{equation}
Both algebras are parabolic, with $\mf{so}(16)_\pm$ as their Levi subalgebras.
In the supergravity realisation $\mf{q}_\pm$ correspond to chiral and anti-chiral
subalgebras \cite{NS} (as we will make explicit below). 
\medskip

\noindent
{\bf Proof:}
We only need to show the subalgebras $\mf{q}_\pm\subset \mf{q}$ commute since this will imply their direct sum structures. In order to see that they commute we take recourse to~\eqref{eq:qpm2} which tells us that we can find representatives of $\mf{q}_\pm$ of the form $x_\mp + \mf{i}$ with $x_\mp\in\mf{i}_\mp$. Then the commutator is
\begin{align}
\lb x_- +\mf{i} ,\, x_+ +\mf{i} \rb \in \mf{i}_+ \cap \mf{i}_-  \,,
\end{align}
which is equal to zero in $\mf{q}$. \hfill \qedsymbol

\medskip
These observations imply that {\em for all finite-dimensional} $K(\mf{e}_9)$ representations
the action of the algebra splits into a direct sum of mutually commuting chiral and anti-chiral
parabolic algebras. Hence we conclude that the quotient algebra ({\em alias} 
the generalised holonomy algebra)  has the form
\beq\label{GH}
\mf{q}\,=\, \big( \mf{so}(16)_+ \oplus \mf{n}_+ \big) \oplus \big(\mf{so}(16)_- \oplus \mf{n}_- \big)
\eeq
where the nilpotent algebras $\mf{n}_\pm$ are $N$-step nilpotent.
We repeat that this structure holds {\em only} 
for the finite-dimensional representations considered here, but is not true generally 
because neither $\mf{so}(16)_+\oplus \mf{so}(16)_-$ nor its parabolic extensions
are subalgebras of $K(\mf{e}_9)$. They only arise as quotients via the above construction and depend on the choice of ideal through a representation. The group-theoretic version
of \eqref{GH} is encapsulated in \eqref{GHG} of the introduction.
We note that under the exponential map one obtains $Spin(16)$ or $SO(16)$ realised faithfully, depending on whether the starting representation $\tV_0$ is a representation of $SO(16)$ or its spin cover.

\section{Examples and relation to root basis form}
\label{sec:rtbas}

In this section we exemplify the abstract considerations above in terms of concretely 
known representations, referred to as spin-$\frac12$, spin-$\frac32$
spin-$\frac52$ and spin-$\frac72$.
These are actually representations of $K(E_{10})$~\cite{KN3,KN5,KNV}, but we will here focus on
the affine subalgebra $K(E_9)$ by restricting the $E_{10}$ roots to the $K(E_9)$ 
root system (see \cite{KNP} for a more detailed analysis of this embedding).
With the $E_8$ roots $\bal,\bbe,...$ and the affine null root $\de$, the 
roots of affine $E_9$ are $m\de$ (null roots) and $m\de +\bal$ (real roots)
for $m\in\ints$. The $E_9$ generators are thus
\beq
E(m\de + \bal) \quad\mbox{and} \quad H^i(m\de) \quad \mbox{for $m\in\ints$}
\eeq
plus the central charge, which will however drop out.
The corresponding $K(E_9)$ generators are
\bea
J(m\de +\bal) \,&:=&\,    E(m\de +\bal) - E(-m\de - \bal) \quad 
\quad \mbox{for $m\geq 0$}     \nonumber\\[2mm]
J^i(m\de) \,&:=&\,  H^i(m\de) - H^i(-m\de) \quad\qquad \mbox{for $m\geq 1$}
\eea
so that 
\beq\label{J}
J(\ga) = - J(-\ga)
\eeq
for all $E_9$ roots $\ga$.

\subsection{\texorpdfstring{Spin-$\frac12$}{Spin-1/2}}

Recall that the generators of $\widetilde{K(E_8)} \equiv$ Spin(16) can be expressed by the gamma matrices
$\Ga(\bal)$, where $\bal$ runs over the {\em positive} roots of $E_8$ (see \cite{KN3,KN5,KNV}
for details). These descend from
the real 32-by-32 gamma matrices $\Ga(\al)$ for $K(E_{10})$ (where $\al$ is a root
of $E_{10}$), but decompose into two blocks of
16-by-16 matrices for $K(E_8)$. Furthermore we have the mod 2 property
(also valid for $K(E_{10})$)
\beq
\Ga(\bal) = \Ga(-\bal) = \Ga(\bal + 2n\bbe)  \qquad \forall n\in \ints
\eeq
for any elements $\bal$ and $\bbe$ of the  $E_8$ root lattice. To ensure \eqref{J} 
one would have to include an extra co-boundary factor
\beq
\tG(\bal) \equiv \si_{\bal} \Ga(\bal)  \quad , \qquad \si_{\bal} \si_{-\bal} = -1
\eeq
but this subtlety can be ignored as long as we are dealing only with positive roots
(as will be the case in the remainder).

The general form of real $K(E_{10})$ generators in this language for the representations found in~{\cite{KN3,KN5,KNV} is then
\begin{align}
\label{eq:HSKE10}
J(\alpha) = -2\, \rX(\alpha) \otimes \Gamma(\alpha)\,,
\end{align}
where  the `polarisation tensor' $\rX(\alpha)$ is constructed only out of data associated with the real root~$\alpha$. For spin-$\tfrac12$, the tensor $\rX(\alpha)=-\tfrac14$ is a pure number.

For $K(E_9)$ we need the null root $\de$ as an extra root in addition to the
$E_8$ roots. From the gamma 
matrix relations derived before we have (with the wall basis of \cite{KN3,KN5})
\beq
\Ga(\de) \,\equiv\,  \Ga^2 \cdots \Ga^{10} \,=\, \Ga^0 \Ga^1
\eeq
which is just the two-dimensional chirality matrix (note that $\Ga(2m\de) \equiv {\bf{1}}$).
Hence, in this representation, $K(E_9)$ decomposes into mutually commuting chiral 
and anti-chiral Spin(16) algebras, with generators (recall that $\al\cdot\de = 0$ for all
$E_9$ roots)
\beq
\frac12 \big( {\bf{1}} \pm \Ga(\de)\big)  \Ga(\bal) \equiv
\frac12 \big( \Ga(\bal) \pm \Ga(\bal+\de)\big) \qquad (\bal > 0)
\eeq
In terms of the previous abstract construction
the chiral doubling is already apparent from the sign ambiguity in
(\ref{rhoXY}) where the positive (negative) chirality gets tied to the
two fixed points of the involution in the spectral parameter plane \cite{NS}.

\subsection{\texorpdfstring{Spin-$\frac32$}{Spin-3/2}}

From \cite{KN3} we recall the correspondence with the generators written in spin-$\frac32$ form
\bea\label{XGa}
J(m\de + \bal)\, &\cong& \, \rX_{\Ta\Tb}(m\de+\bal) \Ga(m\de + \bal) \nonumber\\[2mm]
J^i(m\de) \,&\cong&\, m\,\xi^i_{[\Ta} \de_{\Tb]} \Ga(m\de)
\eea
where the symmetric 10$\,\times\,$10 matrices $\rX_{\Ta\Tb}$ are defined by
\begin{equation}\label{Xab}
\rX_{\Ta\Tb}(\al) \, :=\,  -\frac12 \al_{\Ta} \al_{\Tb} + \frac14 G_{\Ta\Tb}
\,=\, \rX_{\Ta\Tb}(-\al)
\end{equation}
with the Lorentzian metric $G_{\Ta\Tb} \equiv  1 - \de_{\Ta\Tb}$, and for any real $E_9$ root $\al$
(as we said this formula is actually valid also for $E_{10}$).

For the further correspondence with the generators $\cX^{IJ}_m$ and $\cY^A_m$
we now take $\bal>0$ to be a positive $E_8$ root, and identify\footnote{There is no need
 to spell out the explicit relation between the generators $\cX_m^{IJ}$ and $\cY_m^A$
 and the root basis here, as we hope the notation is self-explanatory.} 
\bea\label{XmYm}
\cX_m (\bal)\,&=&\, \frac12 \Big[ \rX(m\de + \bal) + \rX(m\de - \bal) \Big] \Ga(m\de +\bal) 
    \qquad (\bal >0\,,\, m\geq 0) \nonumber\\[2mm]
 \cY_m(\bal) \,&=&\, \frac12 \Big[ \rX(m\de + \bal) - \rX(m\de - \bal) \Big] \Ga(m\de +\bal) 
 \qquad (\bal >0\,,\, m\geq 1) \nonumber\\[2mm]
 && \qquad \mbox{and} \quad m (\xi^i\wedge \de) \, \Ga(m\de) \qquad
    \mbox{(for $m\geq 1$)}
 \eea
To determine the ideal relations directly from these formulas we note 
the elementary identities which follow directly from \eqref{Xab}
\bea
\frac12 \Big[ \rX(n\de + \bal) + \rX(n\de - \bal) \Big] \,&=&\,
n^2 \cdot \frac12 \Big[ \rX(\de + \bal) + \rX(\de - \bal) \Big]  + (1-n^2) \rX(\bal) \,,\nonumber\\[2mm]
\frac12 \Big[ \rX(n\de + \bal) - \rX(n\de - \bal) \Big] \,&=&\,
 n \cdot \frac12 \Big[ \rX(\de + \bal) - \rX(\de - \bal) \Big] \,.
 \eea
Here, we already recognise some of the previously derived relations, but we 
still need to keep track of the extra $\Ga$-matrix factors in \eqref{XmYm}
which differ according to whether $\de$ is multiplied by an even or an odd integer. 
Not mixing even and odd multiples of $\de$, we thus derive the relations
\begin{align}
\label{X2n}
\cX_{2n} &= n^2 \cX_2 + (1-n^2) \cX_0 \,,\nonumber\\[2mm]
\cX_{2n+1} &= \frac18 [(2n+1)^2 -1]\cX_3 -\frac18[(2n+1)^2 -9] \cX_1 \,,\nonumber\\[2mm]
\cY_{2n} &= n \cY_2 \quad , \qquad \cY_{2n+1} = (2n+1) \cY_1\,.
\end{align}
For $n\geq 2$ these relations correspond precisely to the combinations in 
\eqref{ieven} which span the intersection ideal $\mf{i}_+\cap \mf{i}_-$.
As we saw above, in the parabolic basis this corresponds to the truncation
\beq
\tA_4 = \tA_6 = \cdots = 0 \quad ,\qquad
\tS_3 = \tS_5 = \cdots = 0\,.
\eeq

To get the chiral combinations we must now combine the $\rX$ factors with the `wrong'
$\Ga$-matrices by means of these relations, and re-express them 
in terms of $\cX$ generators; idem for $\cY$.
For instance,
\beq
 \frac12 \Big[ \rX(\de + \bal) + \rX(\de - \bal) \Big] \Ga(\bal) \,=\,
 \frac14 \left( \frac12 \Big[ \rX(2\de + \bal) + \rX(2\de - \bal) \Big] 
    + 3 \rX(\bal)\right) \Ga(\bal)
\eeq
from which obtain the chiral combinations
\beq
 \frac12 \Big[ \rX(\de + \bal) + \rX(\de - \bal) \Big] \Ga(\de + \bal)
 \big({\bf{1}} \pm \Ga(\de)\big) 
\,=\, \cX_1 \pm \frac14\big(\cX_2 + 3\cX_0 \big)
\eeq
which are, respectively, elements of $\mf{i}_+$ and $\mf{i}_-$.
Similarly,
\beq
\rX(\bal) \Ga(\bal) \big( {\bf{1}} \pm \Ga(\de)\big) = \cX_0 \mp \frac18 \big(\cX_3 - 9\cX_1\big)\,.
\eeq
For the $\cY$ generators the relevant relations are even simpler: we have
\beq
\frac12 \Big[ \rX(\de + \bal) - \rX(\de - \bal) \Big] \Ga(\de +\bal) 
 \big({\bf{1}} \pm \Ga(\de)\big) \,=\, \cY_1 \pm \frac12 \cY_2\,.
 \eeq
Hence, all chiral and ant-chiral generators can be expressed as linear combinations of 
$\cX_0$, $\cX_1$, $\cX_2$, $\cX_3$ and $\cY_1$, $\cY_2$, giving a total of  
$2\times (120 + 128 + 120)$ generators, which is the correct count
corresponding to $\mf{k}/\mf{i}_+\cap \mf{i}_-$, and in complete agreement with the
supergravity analysis \cite{NS}.

To check that chiral and anti-chiral transformations commute, we compute (for instance)
\begin{align}
\Big[ \cY_1 + \frac12 \cY_2\,,\, \cY_1 - \frac12 \cY_2 \Big] \,&=\,
\frac14 \big( - \cX_4 + 4\cX_2 - 3\cX_0 \big) \nonumber\,,\\[2mm]
\Big[ \cX_1 + \frac14\big(\cX_2 + 3\cX_0 \big) \,,\, 
\cX_1  - \frac14\big(\cX_2 + 3\cX_0 \big) \Big] \,&=\,  \frac1{16}(-\cX_4 + 4\cX_2 - 3\cX_0) \,.
\end{align}
Both commutators vanish by the ideal relation \eqref{X2n} (actually identically in
this explicit representation!), so chiral and anti-chiral
transformations indeed commute by virtue of the ideal relations. In the
explicit representation with chiral projectors this commutation property is 
of course obvious, and merely confirms the abstract argument given before. 

As shown in the preceding section one can therefore abstract from such concrete realisations,
and define the commuting chiral algebras entirely in terms of the $\cX_n$ and $\cY_n$
and ideal relations, such that closure of the given subalgebra depends on the respective 
ideal relations.

\subsection{\texorpdfstring{Spin-$\frac52$}{Spin-5/2}}

For spin-$\frac52$ we have a similar representation as in (\ref{XGa}), except that the 
matrices $\rX_{\Ta\Tb|\Tc\Td}$ now act in a $55$-dimensional space \cite{KN3}, where
\begin{equation}\label{Xabcd}
\rX_{\Ta\Tb,\Tc\Td}(\al)  \,\coloneqq\, \frac12 \al_\Ta \al_\Tb \al_\Tc \al_\Td -
 \al_{(\Ta} G_{\Tb)(\Tc} \al_{\Td)} + \frac14 G_{\Ta(\Tc} G_{\Td)\Tb}\,.
\end{equation}
The formula \eqref{Xabcd} now yields the relations 
\bea
\frac12 \Big[ \rX(n\de + \bal) + \rX(n\de - \bal) \Big] \,&=&\,
\frac1{12} n^2(n^2 -1) \cdot \frac12 \Big[ \rX(2\de + \bal) + \rX(2\de + \bal) \Big] \\[2mm]
&&  \hspace{-1.9cm}
  - \frac13n^2(n^2-4) \cdot \frac12 \Big[ \rX(\de + \bal) + \rX(\de - \bal) \Big] 
+ \frac14(n^4 - 5n^2 +4) \rX(\bal)   \nonumber
\eea
and
\begin{align}
 \frac12 \Big[ \rX(n\de + \bal) - \rX(n\de - \bal) \Big] \,&=\,
\frac1{24}(n^3- n) \cdot \frac12 \Big[ \rX(3\de + \bal) - \rX(3\de - \bal) \Big] \nonumber\\[2mm]
&\quad\quad + \frac18 (-n^3 +9n)  \cdot \frac12 \Big[ \rX(\de + \bal) - \rX(\de - \bal) \Big] \,.
\end{align}
These are exactly the relations we have found above in~\eqref{eq:ideal1} with $\K=2$, i.e., for
\beq
\tA_6 = \tA_8 = \cdots = 0 \quad ,\qquad
\tS_5 = \tS_7 = \cdots = 0\,.
\eeq

\subsection{\texorpdfstring{Spin-$\frac72$}{Spin-7/2}}

For spin-$\tfrac72$ the polarisation matrices $\rX_{\Ta_1\Ta_2\Ta_3|\Tb_1\Tb_2\Tb_3}(\alpha)=\rX_{(\Ta_1\Ta_2\Ta_3)|(\Tb_1\Tb_2\Tb_3)}(\alpha)$ are given by~\cite{KN3}
\begin{align}
\label{eq:X72}
X(\alpha)^{\Ta_1\Ta_2\Ta_3}{}_{\Tb_1\Tb_2\Tb_3}& =-\frac13 \alpha^{\Ta_1} \alpha^{\Ta_2}\alpha^{\Ta_3} \alpha_{\Tb_1}\alpha_{\Tb_2}\alpha_{\Tb_3}
+\frac32 \alpha^{(\Ta_1} \alpha^{\Ta_2}\delta^{\Ta_3)}_{(\Tb_1} \alpha_{\Tb_2}\alpha_{\Tb_3)}
-\frac32\alpha^{(\Ta_1}\delta^{\Ta_2}_{(\Tb_1} \delta^{\Ta_3)}_{\Tb_2\phantom{)}}\alpha_{\Tb_3)}^{\phantom{)}}&\nn\\
&\quad +\frac14 \delta^{(\Ta_1}_{(\Tb_1}\delta^{\Ta_2\phantom{)}}_{\Tb_2\phantom{)}}\delta^{\Ta_3)}_{\Tb_3)}
+\frac1{12}(2-\sqrt{3}) \alpha^{(\Ta_1}  G^{\Ta_2\Ta_3)}G_{(\Tb_1\Tb_2}\alpha_{\Tb_3)}&\\
&\quad+\frac1{12}(-1+\sqrt{3})\left( \alpha^{(\Ta_1}\alpha^{\Ta_2}\alpha^{\Ta_3)}G_{(\Tb_1\Tb_2}\alpha_{\Tb_3)}
+\alpha^{(\Ta_1}  G^{\Ta_2\Ta_3)}\alpha_{(\Tb_1}\alpha_{\Tb_2}\alpha_{\Tb_3)}\right)\,,\nn
\end{align} 
where as before we raise (and lower) indices with the DeWitt metric $G_{\Ta\Tb}$. 
Restricting from $E_{10}$ roots $\al$ to affine roots $m\de \pm \bal$
one can deduce the following relations from~\eqref{eq:X72} 
\begin{align}
&\quad \frac12 \Big[ \rX_{\Ta_1\Ta_2\Ta_3|\Tb_1\Tb_2\Tb_3}(n\delta + \bal) + 
\rX_{\Ta_1\Ta_2\Ta_3|\Tb_1\Tb_2\Tb_3}(n\delta - \bal)\Big] \nn\\
&=
 -\frac1{36} (n^2-9)(n^2-4)(n^2-1) \rX_{\Ta_1\Ta_2\Ta_3|\Tb_1\Tb_2\Tb_3}( \bal)\nn\\
&\quad  +\frac1{48} (n^2-9)(n^2-4)n^2\Big[ \rX_{\Ta_1\Ta_2\Ta_3|\Tb_1\Tb_2\Tb_3}(\delta + \bal) + \rX_{\Ta_1\Ta_2\Ta_3|\Tb_1\Tb_2\Tb_3}(\delta - \bal)\Big]\nn\\
&\quad   -\frac1{120} (n^2-9)(n^2-1)n^2 \Big[ \rX_{\Ta_1\Ta_2\Ta_3|\Tb_1\Tb_2\Tb_3}(2\delta + \bal) + \rX_{\Ta_1\Ta_2\Ta_3|\Tb_1\Tb_2\Tb_3}(2\delta - \bal)\Big]\nn\\
&\quad +\frac1{720} (n^2-4)(n^2-1)n^2 \Big[ \rX_{\Ta_1\Ta_2\Ta_3|\Tb_1\Tb_2\Tb_3}(3\delta + \bal) + \rX_{\Ta_1\Ta_2\Ta_3|\Tb_1\Tb_2\Tb_3}(3\delta - \bal)\Big]
\end{align}
and
\begin{align}
&\quad \frac12 \Big[ \rX_{\Ta_1\Ta_2\Ta_3|\Tb_1\Tb_2\Tb_3}(n\delta + \bal) - \rX_{\Ta_1\Ta_2\Ta_3|\Tb_1\Tb_2\Tb_3}(n\delta - \bal)\Big] \nn\\
&=
 \frac1{48} (n^2-9)(n^2-4)n\Big[ \rX_{\Ta_1\Ta_2\Ta_3|\Tb_1\Tb_2\Tb_3}(\delta + \bal) - \rX_{\Ta_1\Ta_2\Ta_3|\Tb_1\Tb_2\Tb_3}(\delta - \bal)\Big]\nn\\
&\quad   -\frac1{60} (n^2-9)(n^2-1)n \Big[ \rX_{\Ta_1\Ta_2\Ta_3|\Tb_1\Tb_2\Tb_3}(2\delta + \bal) - \rX_{\Ta_1\Ta_2\Ta_3|\Tb_1\Tb_2\Tb_3}(2\delta - \bal)\Big]\nn\\
&\quad +\frac1{240} (n^2-4)(n^2-1)n \Big[ \rX_{\Ta_1\Ta_2\Ta_3|\Tb_1\Tb_2\Tb_3}(3\delta + \bal) - \rX_{\Ta_1\Ta_2\Ta_3|\Tb_1\Tb_2\Tb_3}(3\delta - \bal)\Big]\,.
\end{align}
These correspond precisely to the relations obtained  by truncation with $\K=3$ in~\eqref{eq:ideal1}. Remarkably, the square root appearing in~\eqref{eq:X72} that is needed for satisfying the 
Berman relations does not affect the ideal relations. The appearance of `strange' prefactors 
in \eqref{eq:X72} (which are absolutely necessary for the $K(E_{10})$ Berman relations
to work, see following section) is another clear indication of subtleties that go beyond 
the affine construction.

\section{\texorpdfstring{Extension to $K(\mf{e}_{10})$}{Extension to K(E10)}}
\label{sec:E10ext}

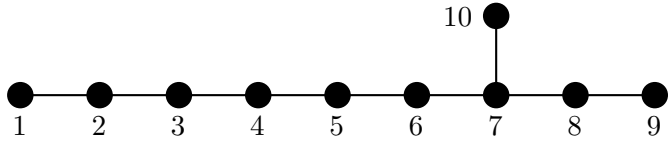
\begin{figure}[t!]
\centering
\begin{picture}(300,50)
\thicklines
\multiput(10,10)(30,0){9}{\circle*{10}}
\put(10,10){\line(1,0){240}}
\put(190,40){\circle*{10}}
\put(190,10){\line(0,1){30}}
\put(7,-5){$1$}
\put(37,-5){$2$}
\put(67,-5){$3$}
\put(97,-5){$4$}
\put(127,-5){$5$}
\put(157,-5){$6$}
\put(187,-5){$7$}
\put(217,-5){$8$}
\put(247,-5){$9$}
\put(170,36){$10$}
\end{picture}
\caption{\label{fig:e10dynk}{\sl Dynkin diagram of $E_{10}$ with nodes labelled.}}
\end{figure}

As we already mentioned the ultimate goal of the present investigation
is a better understanding of the representation theory for the maximally hyperbolic
algebra $\mf{e}_{10}$ and its maximal compact subalgebra $K(\mf{e}_{10})$. The
$E_{10}/K(E_{10})$ model of \cite{DHN} is conjectured to provide a framework for 
(the bosonic sector of) M-theory in a small tension limit, with (de-)emergent space 
at the singularity, such that a precise correspondence with the field 
content and equations of motion of $D=11$ supergravity holds if the
field theory is restricted to first order spatial gradients, and the Lie algebra
expansion to levels $\ell\leq 3$ (or roots of height ht$(\alpha) \leq 29$) in a decomposition
of $\mf{e}_{10}$ under its $\mf{gl}_{10}$ subalgebra. Consequently the main challenge 
that remains is to extend this correspondence to higher order spatial gradients and the 
exponentially growing tower of $\ell > 3$ representations, and this is a problem that 
has defied solution until now. It is here that the study of fermions could
make a crucial difference, because the existence of unfaithful $K(\mf{e}_{10})$
representations of ever increasing dimensions may afford a more `global' view of $\widetilde{K(E_{10})}$,
and thus also $E_{10}$, which is unachievable with the tools usually employed
in the theory of Kac--Moody algebras and Kac--Moody groups (partition functions, level 
decompositions, local constructions, {\em etc.}). This final section is meant to provide
a sketchy account of how progress on the representation theory of  $K(\mf{e}_9)$ could 
be possibly exploited towards a better understanding of $K(\mf{e}_{10})$. Namely, the available 
evidence indicates that this can be accomplished by `gluing' chiral and anti-chiral
representations of $K(\mf{e}_9)$ by means of an explicit action of the  one 
remaining Berman generator $x_1$ on the given representations. However, the 
consistency conditions that must be satisfied (cf. \eqref{eq:B12}, \eqref{eq:B1} and 
\eqref{eq:B2} below) remain to be explored.

In order to discuss the extension from $K(\mf{e}_9)$ to $K(\mf{e}_{10})$ we make use of the labelling of the Dynkin diagram shown in figure~\ref{fig:e10dynk}. The algebra $K(\mf{e}_{10})$ is generated by ten Berman generators $x_1,\ldots,x_{10}$ with Berman relations~\cite{B}
\begin{align}
\lb x_i, x_j \rb &=0 &&\text{if $i$ and $j$ disconnected in the diagram,}\nn\\
\lb x_i \lb x_i, x_j \rb\rb  &=-x_j &&\text{if $i$ and $j$ connected in the diagram.}
\end{align}
The $K(\mf{e}_9)$ subalgebra is obtained by restricting to the abstract generators $x_2,\ldots, x_{10}$.

For any finite-dimensional representation of $K(\mf{e}_9)$, the Berman generators $x_2,\ldots, x_{10}$ are represented by specific matrices $X_2,\ldots, X_{10}$ that can be written as 
\begin{equation}
X_i \,:=\, 
\frac14 \left(\gamma_{\alpha\beta}^{i,i+1} \cX_0^{\alpha\beta} + 
\gamma^{i,i+1}_{\dot\alpha\dot\beta} \cX_0^{\dot\alpha\dot\beta}\right) 
\end{equation}
for $i=3,...,9$, and
\begin{equation}
X_{10} \,:=\, - \frac12 \gamma^{8\,9\,10}_{\alpha\dot\beta} \,\cX_0^{\alpha\dot\beta} \quad \mbox{and} \qquad
X_2 \,:=\, - \frac12 \gamma^2_{\alpha\dot\beta} \left( \cX_1^{\alpha\dot\beta} + \cY_1^{\alpha\dot\beta} \right)\,,
\end{equation}
where all $\cX_n$ and $\cY_n$ are realised as {\em finite-dimensional} matrices in
the concrete unfaithful representation under consideration. In the above equations we 
use the notation of~\cite{KNP} as well as the decomposition of the $SO(16)$ indices
$[IJ]$ and $A$ under the $SO(8)\times SO(8)$ subgroup of $SO(16)$ explained there.

We then seek a realisation of the $K(\mf{e}_{10})$ Berman generators for the
`doubled' representation $S_+\oplus S_-$ by means of the ansatz
\begin{equation}
x_1 = \begin{pmatrix} 0 & B \\ C& 0 \end{pmatrix} \;\; , \quad
x_2 = \begin{pmatrix} X_2 & 0 \\ 0& X_2 \end{pmatrix} \;\;,\quad
x_i = \begin{pmatrix} X_i & 0 \\ 0 & X_i \end{pmatrix} \;\;,\quad
x_{10} = \begin{pmatrix} X_{10}  & 0 \\ 0 & - X_{10} \end{pmatrix} 
\end{equation}
with $i= 3,...,9$ as before.  The relation $[x_1, x_j] =0$ for $j=3,...,10$ requires
\begin{equation}
[B , X_j] = [C , X_j ] = 0
\end{equation}
or $j=3,\ldots,9$, and $\{ B,X_{10}\}= \{C,X_{10}\}=0$. Therefore, 
$B$ and $C$ must be SO(16) singlets. Since the Berman relations for $j=2,...,10$
are satisfied by construction, the critical Berman
relations are the two remaining ones which read
\begin{equation}
\label{eq:B12}
[x_1, [x_1, x_2]] = - x_2 \quad , \qquad
[x_2,[x_2 , x_1]] = - x_1
\end{equation}
These follow if 
\begin{equation} 
\label{eq:B1}
\{ BC \,,\, X_2\}  - 2 BX_2 C  \,\stackrel{!}{=}\, - X_2 \;\;, \quad
\{ CB \,,\, X_2\}  - 2 CX_2 B \, \stackrel{!}{=} \,- X_2
\end{equation}
and
\begin{equation}
\label{eq:B2}
\lb X_2, \lb X_2, B \rb\rb \stackrel{!}{=} -B\,,\quad 
\lb X_2, \lb X_2, C \rb\rb \stackrel{!}{=} - C\,.
\end{equation}

If $x_1$ simply rotated the two chiralities, we would obtain that $B=-C=\tfrac12$. While this solves~\eqref{eq:B1}, 
it clearly does not solve~\eqref{eq:B2} and therefore this is not a viable solution. From the physics 
perspective this is clear, because the Berman generator $x_1$ is associated to a spatial $\mf{so}(2)$ rotation, 
whereas the chirality refers to the behaviour under a non-compact space-time $\mf{so}(1,1)$
(the group $SO(1,1)$ scales the chiral and antichiral parts with inverse factors). 

This is borne out 
in the example of spin-$\frac12$ where one check that the equations~\eqref{eq:B1} 
and~\eqref{eq:B2} are indeed satisfied with 
$B=-C = -\frac12 \gamma^2$ and $X_2 = \frac12 \gamma^{23}$.
For the higher spin examples $s= \frac32, \frac52, \frac72$ we can similarly read off the solution 
from the explicit formulas of the foregoing section, with the result
\beq\label{BC}
B \,=\,  - C \,=\, -2\, \rX(\al_1) \otimes \ga^2
\eeq
where $\al_1$ is the over-extended (left-most) simple root in the $E_{10}$ Dynkin diagram. This formula is valid for all representations of $K(\mf{e}_{10})$ constructed as in~\eqref{eq:HSKE10}. The relations~\eqref{eq:B1} and~\eqref{eq:B2} are then implied by the properties of the tensor $\rX(\alpha_1)$ together with $X_2=-2 \,\rX(\alpha_2) \otimes \gamma^{23}$.

For any given unfaithful representation of $K(E_9)$ the problem of uplifting
it to an unfaithful representation of $K(E_{10})$ is thus reduced to finding
finite-dimensional matrices $B$ and $C$ solving the relations \eqref{eq:B1} and \eqref{eq:B2}. 
Consider for example the (chiral) $K(\mf{e}_9)$ representation 
that consists of $\tV_0\cong {\bf 16}$ and $\tV_1\cong {\bf 128}_c$. Doubling this chiral 
representation leads to a $288$-dimensional space that can {\em not} be turned into 
a $K(\mf{e}_{10})$ representation, {\em i.e.}, no $x_1$ satisfying the relations~\eqref{eq:B12} exists,
because the spin-$\tfrac32$ representation of $K(\mf{e}_{10})$ requires 
a second $\tV_2\cong {\bf 16}$ in the chiral $K(\mf{e}_9)$ representation~\cite{KNP}.
We therefore see from the explicit examples that it is by no means trivial that matrices 
$B$ and $C$ exist, and it is furthermore quite possible that a more general
ansatz than \eqref{BC} may be needed.



\begin{thebibliography}{32}
\setlength{\itemsep}{0pt}

\bibitem{Kleinschmidt:2021agj}
A.~Kleinschmidt, R.~K\"ohl, R.~Lautenbacher and H.~Nicolai,
``Representations of involutory subalgebras of affine Kac-Moody algebras,''
\doi{Commun. Math. Phys. \textbf{392} (2022) no.1, 89-123}{doi:10.1007/s00220-022-04342-9}
\eprintRT{2102.00870}.


\bibitem{CJ}
E.~Cremmer and B.~Julia,
``The SO(8) Supergravity,''
\doi{Nucl. Phys. B \textbf{159} (1979), 141-212}{doi:10.1016/0550-3213(79)90331-6}.

\bibitem{dWN} 
B.~de Wit and H.~Nicolai,
``$d=11$ Supergravity With Local SU(8) Invariance,''
\doi{Nucl. Phys. B \textbf{274} (1986), 363-400}{doi:10.1016/0550-3213(86)90290-7}.

\bibitem{Duff:1990xz}
M.~J.~Duff and K.~S.~Stelle,
``Multimembrane solutions of D = 11 supergravity,''
\doi{Phys. Lett. B \textbf{253} (1991), 113-118}{doi:10.1016/0370-2693(91)91371-2}.

\bibitem{Duff:2003ec}
M.~J.~Duff and J.~T.~Liu,
``Hidden space-time symmetries and generalized holonomy in M theory,''
\doi{Nucl. Phys. B \textbf{674} (2003), 217-230}{doi:10.1016/j.nuclphysb.2003.09.019}
\eprint{hep-th/0303140}.

\bibitem{Hull:2003mf}
C.~Hull,
``Holonomy and symmetry in M theory,''
\eprint{hep-th/0305039}.

\bibitem{Lu:2005im}
H.~Lu, C.~N.~Pope and K.~S.~Stelle,
``Generalised holonomy for higher-order corrections to supersymmetric backgrounds in string and M-theory,''
\doi{Nucl. Phys. B \textbf{741} (2006), 17-33}{doi:10.1016/j.nuclphysb.2006.01.042}
\eprint{hep-th/0509057}.

\bibitem{Grana:2005jc}
M.~Gra\~na,
``Flux compactifications in string theory: A Comprehensive review,''
\doi{Phys. Rept. \textbf{423} (2006), 91-158}{doi:10.1016/j.physrep.2005.10.008}
\eprint{hep-th/0509003}.

\bibitem{Grana:2005sn}
M.~Gra\~na, R.~Minasian, M.~Petrini and A.~Tomasiello,
``Generalized structures of N=1 vacua,''
\doi{JHEP \textbf{11} (2005), 020}{doi:10.1088/1126-6708/2005/11/020}
\eprint{hep-th/0505212}.

\bibitem{Gabella:2009wu}
M.~Gabella, J.~P.~Gauntlett, E.~Palti, J.~Sparks and D.~Waldram,
``AdS(5) Solutions of Type IIB Supergravity and Generalized Complex Geometry,''
\doi{Commun. Math. Phys. \textbf{299} (2010), 365-408}{doi:10.1007/s00220-010-1083-y}
\eprintN{0906.4109}.

\bibitem{Coimbra:2014uxa}
A.~Coimbra, C.~Strickland-Constable and D.~Waldram,
``Supersymmetric Backgrounds and Generalised Special Holonomy,''
\doi{Class. Quant. Grav. \textbf{33} (2016) no.12, 125026}{doi:10.1088/0264-9381/33/12/125026}
\eprintN{1411.5721}.

\bibitem{Coimbra:2016ydd}
A.~Coimbra and C.~Strickland-Constable,
``Supersymmetric Backgrounds, the Killing Superalgebra, and Generalised Special Holonomy,''
\doi{JHEP \textbf{11} (2016), 063}{doi:10.1007/JHEP11(2016)063}
\eprintN{1606.09304}.

\bibitem{West:2003fc}
P.~C.~West,
``E(11), SL(32) and central charges,''
\doi{Phys. Lett. B \textbf{575} (2003), 333-342}{doi:10.1016/j.physletb.2003.09.059}
\eprint{hep-th/0307098}.

\bibitem{Keurentjes:2003yu}
A.~Keurentjes,
``The Topology of U duality (sub)groups,''
\doi{Class. Quant. Grav. \textbf{21} (2004) no.6, 1695-1708}{doi:10.1088/0264-9381/21/6/025}
\eprint{hep-th/0309106}.


\bibitem{deBuyl:2005zy}
S.~de Buyl, M.~Henneaux and L.~Paulot,
``Hidden symmetries and Dirac fermions,''
\doi{Class. Quant. Grav. \textbf{22} (2005), 3595-3622}{doi:10.1088/0264-9381/22/17/018}
\eprint{hep-th/0506009}.

\bibitem{Damour:2005zs}
T.~Damour, A.~Kleinschmidt and H.~Nicolai,
``Hidden symmetries and the fermionic sector of eleven-dimensional supergravity,''
\doi{Phys. Lett. B \textbf{634} (2006), 319-324}{doi:10.1016/j.physletb.2006.01.015}
\eprint{hep-th/0512163}.

\bibitem{deBuyl:2005sch}
S.~de Buyl, M.~Henneaux and L.~Paulot,
``Extended $E_8$ invariance of 11-dimensional supergravity,''
\doi{JHEP \textbf{02} (2006), 056}{doi:10.1088/1126-6708/2006/02/056}
\eprint{hep-th/0512292}.

\bibitem{Damour:2006xu}
T.~Damour, A.~Kleinschmidt and H.~Nicolai,
``$K(E_{10})$, Supergravity and Fermions,''
\doi{JHEP \textbf{08} (2006), 046}{doi:10.1088/1126-6708/2006/08/046}
\eprint{hep-th/0606105}.

\bibitem{Harring:2019}
  P.~Harring and R.~K\"ohl,
  ``Fundamental groups of split real Kac--Moody groups and generalized flag manifolds,''
  \eprintGR{1905.13444}.

\bibitem{Kleinschmidt:2006tm}
A.~Kleinschmidt and H.~Nicolai,
``IIA and IIB spinors from $K(E_{10})$,''
\doi{Phys. Lett. B \textbf{637} (2006), 107-112}{doi:10.1016/j.physletb.2006.04.007}
\eprint{hep-th/0603205}.

\bibitem{KNV}
A.~Kleinschmidt, H.~Nicolai and A.~Vigan\`o,
``On spinorial representations of involutory subalgebras of Kac-Moody algebras,''
In: V.~Gritsenko, V.~Spiridonov (eds.) 
\doi{Partition Functions and Automorphic Forms. Moscow Lectures, vol 5. Springer, Cham (2020)}{https://doi.org/10.1007/978-3-030-42400-8_4}
\eprintN{1811.11659}.

\bibitem{Steele:2010tk}
D.~Steele and P.~C.~West,
``E11 and Supersymmetry,''
\doi{JHEP \textbf{02} (2011), 101}{doi:10.1007/JHEP02(2011)101}
\eprintN{1011.5820}.

\bibitem{NS}
H.~Nicolai and H.~Samtleben,
``On $K(E_9)$,''
\doi{Q. J. Pure Appl. Math. \textbf{1} (2005) 180--204}{doi:10.4310/PAMQ.2005.v1.n1.a8}
\eprint{hep-th/0407055}.

\bibitem{Kleinschmidt:2007zd}
A.~Kleinschmidt,
``Unifying R-symmetry in M-theory,''
in: \doi{New trends in Mathematical Physics (Proceedings of the XVth International Congress in Mathematical Physics)}{doi:10.1007/978-90-481-2810-5}, V. Sidoravi\v{c}ius (ed.), Springer (2009)
\eprint{hep-th/0703262}.


\bibitem{KN3}
A.~Kleinschmidt and H.~Nicolai,
``On higher spin realizations of $K(E_{10})$,''
\doi{JHEP \textbf{08} (2013) 041}{doi:10.1007/JHEP08(2013)041}
\eprintN{1307.0413}.

\bibitem{KN5}
A.~Kleinschmidt and H.~Nicolai,
``Higher spin representations of $K(E_{10})$,''
\doi{in: \textit{Higher spin gauge theories}, L. Brink, M. Henneaux, M. Vasiliev (eds.), World Scientific (2017) 25--38}{doi:10.1142/9789813144101\_0003}
\eprintN{1602.04116}.

\bibitem{Julia} B.~Julia, ``Group Disintegrations,'' 
in:
  S.~W.~Hawking and M.~Ro\v{c}ek (eds.), Superspace and
  Supergravity, Proceedings of the Nuffield Workshop,
  Cambridge, Eng., Jun 22 -- Jul 12, 1980, Cambridge
  University Press (Cambridge, 1981) 331--350.
  
\bibitem{Julia2}
  B.~Julia,
  ``Kac--Moody Symmetry of Gravitation
  and Supergravity Theories,''
   in: M.~Flato, P.~Sally
  and G.~Zuckerman (eds.), {\sl Applications of Group Theory in Physics
  and Mathematical Physics} (Lectures in Applied Mathematics {\bf
  21}), Am. Math. Soc. (Providence, 1985) 355--374, LPTENS 82/22.

\bibitem{Nicolai:1987kz}
H.~Nicolai,
``The Integrability of $N=16$ Supergravity,''
\doi{Phys. Lett. B \textbf{194} (1987), 402}{doi:10.1016/0370-2693(87)91072-0}


\bibitem{DHN}
T.~Damour, M.~Henneaux and H.~Nicolai,
``E(10) and a 'small tension expansion' of M theory,''
\doi{Phys. Rev. Lett. \textbf{89} (2002), 221601}{doi:10.1103/PhysRevLett.89.221601}
\eprint{hep-th/0207267}.


\bibitem{Kleinschmidt:2014uwa}
A.~Kleinschmidt, H.~Nicolai and N.~K.~Chidambaram,
``Canonical structure of the E10 model and supersymmetry,''
\doi{Phys. Rev. D \textbf{91} (2015) no.8, 085039}{doi:10.1103/PhysRevD.91.085039}
\eprintN{1411.5893}.

\bibitem{Damour:2009zc}
T.~Damour and C.~Hillmann,
``Fermionic Kac-Moody Billiards and Supergravity,''
\doi{JHEP \textbf{08} (2009) 100}{doi:10.1088/1126-6708/2009/08/100}
\eprintN{0906.3116}.

\bibitem{Hainke:2014}
 G.~Hainke, R.~K\"ohl and  P.~Levy,
 ``Generalized spin representations,'' with an appendix by M.~Horn and R.~K\"ohl,
\doi{M\"unster J. Math. {\bf 8} (2015) 181--210}{10.17879/65219674985}
\eprintRT{1110.5576}.

\bibitem{Lautenbacher:2017}
R.~Lautenbacher and R.~K\"ohl
``Extending generalized spin representations,''
\doi{J. Lie Theory {\bf 28} (2018) 915--940}{https://www.heldermann.de/JLT/JLT28/JLT284/jlt28045.htm}
\eprintRT{1705.00118}.




\bibitem{Hohm:2019bba}
O.~Hohm and H.~Samtleben,
``The many facets of exceptional field theory,''
\doi{PoS \textbf{CORFU2018} (2019), 098}{doi:10.22323/1.347.0098}
\eprintN{1905.08312}.

\bibitem{Berman:2020tqn}
D.~S.~Berman and C.~D.~A.~Blair,
``The Geometry, Branes and Applications of Exceptional Field Theory,''
\doi{Int. J. Mod. Phys. A \textbf{35} (2020) no.30, 2030014}{doi:10.1142/S0217751X20300148}
\eprintN{2006.09777}.



\bibitem{Godazgar:2014nqa}
H.~Godazgar, M.~Godazgar, O.~Hohm, H.~Nicolai and H.~Samtleben,
``Supersymmetric E$_{7(7)}$ Exceptional Field Theory,''
\doi{JHEP \textbf{09} (2014), 044}{doi:10.1007/JHEP09(2014)044}
\eprintN{1406.3235}.

\bibitem{Baguet:2016jph}
A.~Baguet and H.~Samtleben,
``E$_{8(8)}$ Exceptional Field Theory: Geometry, Fermions and Supersymmetry,''
\doi{JHEP \textbf{09} (2016), 168}{doi:10.1007/JHEP09(2016)168}
\eprintN{1607.03119}.

\bibitem{Bossard:2019ksx}
G.~Bossard, A.~Kleinschmidt and E.~Sezgin,
``On supersymmetric E$_{11}$ exceptional field theory,''
\doi{JHEP \textbf{10} (2019), 165}{doi:10.1007/JHEP10(2019)165}
\eprintN{1907.02080}.

\bibitem{Nicolai:1986jk}
H.~Nicolai,
``$D=11$ Supergravity With Local SO(16) Invariance,''
\doi{Phys. Lett. B \textbf{187} (1987), 316-320}{doi:10.1016/0370-2693(87)91102-6}

\bibitem{West:2001as}
P.~C.~West,
``E(11) and M theory,''
\doi{Class. Quant. Grav. \textbf{18} (2001), 4443-4460}{doi:10.1088/0264-9381/18/21/305}
\eprint{hep-th/0104081}.

\bibitem{Bergshoeff:1987cm}
E.~Bergshoeff, E.~Sezgin and P.~K.~Townsend,
``Supermembranes and Eleven-Dimensional Supergravity,''
\doi{Phys. Lett. B \textbf{189} (1987), 75-78}{doi:10.1016/0370-2693(87)91272-X}

\bibitem{deWit:1988wri}
B.~de Wit, J.~Hoppe and H.~Nicolai,
``On the Quantum Mechanics of Supermembranes,''
\doi{Nucl. Phys. B \textbf{305} (1988), 545}{doi:10.1016/0550-3213(88)90116-2}






\bibitem{Breitenlohner:1986um}
P.~Breitenlohner and D.~Maison,
``On the Geroch Group,''
Ann. Inst. H. Poincare Phys. Theor. \textbf{46} (1987), 215.

\bibitem{KNP}
A.~Kleinschmidt, H.~Nicolai and J.~Palmkvist,
``$K(E_9)$ from $K(E_{10})$,''
\doi{JHEP \textbf{06} (2007) 051}{doi:10.1088/1126-6708/2007/06/051}
\eprint{hep-th/0611314}.


\bibitem{B}
  S.~Berman,
 ``On generators and relations for certain involutory subalgebras of Kac-Moody Lie algebras,'' 
\doi{Comm. Algebra {\bf 17} (1989) 3165--3185}{https://doi.org/10.1080/00927878908823899}.

\end{thebibliography}
\end{document}